\documentclass[twocolumn,aps,showpacs]{revtex4}
\usepackage[latin9]{inputenc}
\usepackage{amsmath}
\usepackage{graphicx}
\usepackage{esint}

\makeatletter

\providecommand{\tabularnewline}{\\}

\@ifundefined{textcolor}{}
{%
 \definecolor{BLACK}{gray}{0}
 \definecolor{WHITE}{gray}{1}
 \definecolor{RED}{rgb}{1,0,0}
 \definecolor{GREEN}{rgb}{0,1,0}
 \definecolor{BLUE}{rgb}{0,0,1}
 \definecolor{CYAN}{cmyk}{1,0,0,0}
 \definecolor{MAGENTA}{cmyk}{0,1,0,0}
 \definecolor{YELLOW}{cmyk}{0,0,1,0}
}

\usepackage{color}

\newcommand{\be}{\begin{equation}}
\newcommand{\ee}{  \end{equation}}
\newcommand{\ba}{\begin{eqnarray}}
\newcommand{\ea}{  \end{eqnarray}}

\makeatother

\begin{document}

\title{Defect production in non-equilibrium phase transitions: Experimental
investigation of the Kibble-Zurek mechanism in a two-qubit quantum
simulator}

\author{Jingfu Zhang}
\email{jingfu@e3.physik.tu-dortmund.de}

\affiliation{Fakultät Physik, Technische Universität Dortmund, D-44221 Dortmund,
Germany}

\author{Fernando M. Cucchietti}
\email{fernando@cucchietti.com}

\affiliation{Barcelona Supercomputing Center (BSC), 08034, Barcelona, Spain}

\author{Raymond Laflamme}
\email{laflamme@iqc.ca}

\affiliation{Institute for Quantum Computing, University of Waterloo, Waterloo,
Ontario, N2L 3G1, Canada}

\affiliation{Perimeter Institute for Theoretical Physics, Waterloo, Ontario, N2J
2W9, Canada}

\affiliation{Canadian Institute for Advanced Research, Toronto, Ontario M5G 1Z8,
Canada}

\author{Dieter Suter }
\email{Dieter.Suter@tu-dortmund.de}

\affiliation{Fakultät Physik, Technische Universität Dortmund, D-44221 Dortmund,
Germany }

\date{\today}
\begin{abstract}
Systems passing through quantum critical points at finite rates have
a finite probability of undergoing transitions between different eigenstates
of the instantaneous Hamiltonian. This mechanism was proposed by Kibble
as the underlying mechanism for the formation of topological defects
in the early universe and by Zurek for condensed matter systems. Here,
we use a system of nuclear spins as an experimental quantum simulator
undergoing a non-equilibrium quantum phase transition. The experimental
data confirm the validity of the Kibble-Zurek mechanism of defect
formation.
\end{abstract}

\maketitle

\section{Introduction}

When a system is driven through a continuous phase transition at a
finite rate, domain structures can arise due to spontaneous
symmetry breaking. With the growth of the domains, they can
approach each other and generate, e.g., topological defects. This
process was initially proposed by Kibble for studying the
cosmological phase transition in the early universe
\cite{Kibble,Kibble2}, and developed by Zurek in condensed matter
systems \cite{Zurek,Zurek2,KZMreview}. Today, it is known as the
Kibble-Zurek mechanism (KZM) and has become a universal theory for
studying non-equilibrium dynamics in both classical and quantum
systems \cite{RMP11}.  The KZM owes its appeal and broad
applicability to its universal and scale invariant power law
prediction, based on simple arguments about second-order phase
transitions. Typical examples include critical phenomena in the
quantum Ising model \cite{QIsing1,QIsing2}. The KZM has been
supported by experiments in physical systems, such as superfluids
\cite{superfluid1,superfluid2}, superconductors
\cite{supercond1,supercond2}, Bose-Einstein condensates
\cite{BEC1,BEC2}, colloidal monolayers \cite{kzmcm}, ion crystals
\cite{kzmion1,kzmion2}, and more solid materials
\cite{kzmsm1,kzmsm2}.

 The interest in the KZM was recently enhanced by its
relation with the established Landau-Zener (LZ) model
\cite{Damski05}, where the critical point can be modeled with a
simple avoided level crossing in a two level system. In a true
second-order phase transition, which can only occur in many-body
systems, we would observe instead a symmetry-breaking accompanied
by a closing of the gap. The LZ model cannot be as complex as a
large quantum many-body system with a phase transition, although
we expect to capture at least qualitatively the dynamics of the
change of properties during a quantum phase transition (QPT).
Similar to a second-order QPT, the defect density created during
the passage through the critical point can be controlled through
the quench rate versus the inverse of the gap; if this is large,
it results in a high defect density. In the opposite limit of a
slow quench, the transition through the critical point becomes
almost adiabsatic and the defect density tends to zero. The LZ
model provides a good test-bed for studying the KZM in a
well-controlled system, such as by quantum simulations
\cite{Guo,Guo2,sim3,sim4}.

The essential concept behind the KZM can be introduced as follows.
A thermodynamic system initially in thermal equilibrium evolves under
some control parameter, e.g., pressure or temperature, that drives
a phase transition. If the control parameter changes slowly enough
compared with the relaxation time $\tau$ of the system, the evolution
is adiabatic, i.e., the system can adjust to the new conditions so
that it remains in thermal equilibrium. However, if the control parameter
changes too rapidly, the system cannot adjust to the new conditions
sufficiently fast, and it is driven out of equilibrium. This process
happens mostly near the critical point of the phase transition, where
the relaxation time $\tau$ diverges. In this regime, the dynamics
of the system becomes impulsive and its state is effectively frozen,
i.e., the change of the control parameter in the evolution only introduces
an overall factor to the waveform function of the system. Following
the KZM model \cite{Zurek}, we approximate the evolution as discontinuous
between the two regimes, and define the boundary as the freeze-out
time $\hat{t}$, where the relaxation time of the system is equal
to the time required to reach the critical point at the given (constant)
scan rate.

 Existing experimental investigations of the KZM by
quantum simulators \cite{Guo,Guo2,sim3,sim4} relied on
single-qubit systems. In the case of the Ising model, it is
possible to map the dynamics to the LZ model
\cite{Guo2,sim3,sim4}. In this article, we present an experimental
quantum simulation of the KZM in a system of two interacting
qubits, where the defect generation occurs non-locally,
simultaneously creating quantum entanglement. In our experiment,
we exploited two spin-qubits in a nuclear magnetic resonance (NMR)
system, using the natural Ising interaction of the spins. A small
transverse field was applied to generate the level-anticrossing
for simulating the second order QPT driven by a control field
along $z$-axis. The defect generation and the freezing-out time
were clearly demonstrated. Our implementation in a two-qubit
system can be generalized to larger qubit systems, providing a
test-bed for simulating the KZM in many-body systems.

\section{Theoretical Model}

We first consider the phase diagram of an Ising model consisting of
qubits 1 and 2. Its Hamiltonian is
\begin{equation}
\mathcal{H}^{s}=\sigma_{z}^{1}\sigma_{z}^{2}+B_{z}(\sigma_{z}^{1}+\sigma_{z}^{2}),\label{hamE}
\end{equation}
where $\sigma_{z}^{i}$ denotes the $z$ components of the Pauli operators
and $B_{z}$ the magnetic field component and the coupling strength
between the two qubits has been set to unity. By solving the Hamiltonian
(\ref{hamE}), we obtain the energy levels as $1+2B_{z}$, $-1$,
and $1-2B_{z}$ for the triplet eigenstates $|00\rangle$, $|\phi^{+}\rangle=(|01\rangle+|10\rangle)/\sqrt{2}$,
and $|11\rangle$, respectively. Here, we do not include the singlet
state $|\phi^{-}\rangle=(|01\rangle-|10\rangle)/\sqrt{2}$, since
the symmetry of the Hamiltonian confines the evolution of a triplet
state to the triplet subsystem. Depending on the control parameter
$B_{z}$, the ground state of the system is then
\begin{eqnarray}
|\psi_{g}(B_{z})\rangle= & \left\{ \begin{array}{c}
|00\rangle\\
|\phi^{+}\rangle\\
|11\rangle
\end{array}\right. & \begin{array}{c}
(B_{z}\leq-1)\\
(-1\leq B_{z}\leq1)\\
(B_{z}\geq1)
\end{array}.\label{groundBx0}
\end{eqnarray}
$|\phi^{+}\rangle$ is one of the Bell states, a maximally
entangled state and therefore a useful resource in the field of
quantum information \cite{nielsen,Stolze:2008xy}. As a function of
the control parameter $B_{z}$, the system undergoes a QPT with
critical points $B_{z}=\pm1$ \cite{Peng05,Zhang08}.

\begin{figure}[bt]
\centering{}\includegraphics[width=0.99\columnwidth]{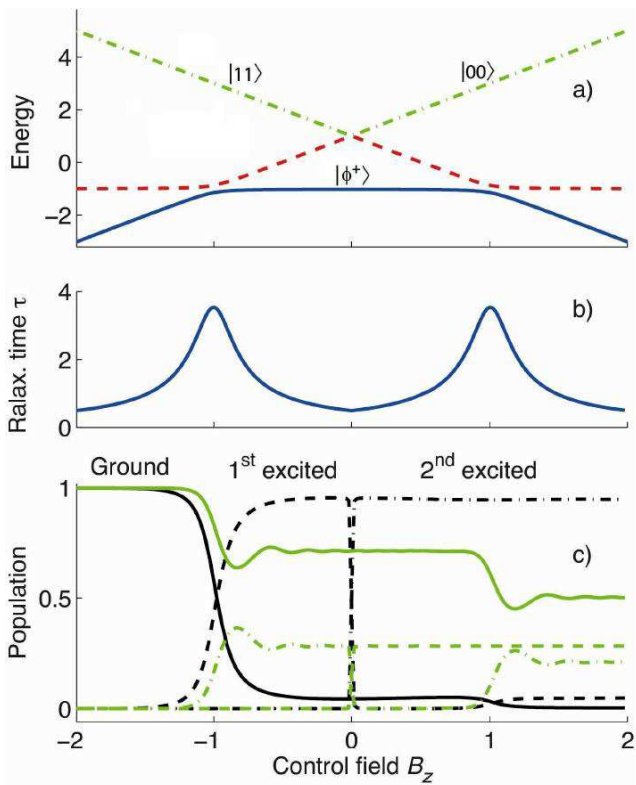}
\caption{Numerical simulation of the two qubit system with
$B_{x}=0.1$. (a) Energy level diagram. The ground, first and
second excited states are indicated by the solid, dashed and
dash-dotted curves, respectively. (b) Relaxation time $\tau$ as a
function of the control parameter. (c) Populations $|a_{i}|^{2}$
during a scan with rates $k=1$ and $1/20$, shown as black (dark)
and green (light) curves, respectively.}

\label{coefficients}
\end{figure}

To simulate the KZM, we drive transitions of the system through the
critical points at $B_{z}=\pm1$. These transitions can be adiabatic
only if there is no exact crossing. These avoided crossings can be
generated by adding a small magnetic field along the $x$-axis. The
Hamiltonian of the system is then
\begin{equation}
\mathcal{H}(t)=B_{x}(\sigma_{x}^{1}+\sigma_{x}^{2})+B_{z}(t)(\sigma_{z}^{1}+\sigma_{z}^{2})+\sigma_{z}^{1}\sigma_{z}^{2}.\label{ham2}
\end{equation}
Figure \ref{coefficients}(a) illustrates the anti-crossing energy
levels for $B_{x}=0.1$. The instantaneous ground state can be represented
as
\begin{equation}
|\psi_{g}(t)\rangle=c_{0}|00\rangle+c_{+}|\phi^{+}\rangle+c_{1}|11\rangle.\label{ground}
\end{equation}
Analytical expressions for the coefficients $c_{0}$, $c_{+}$, and
$c_{1}$ are given, e.g., in \cite{Peng05}.

We initialize the system into its ground state at $t=0$ and let the
field $B_{z}$ change linearly with time,
\begin{equation}
B_{z}(t)=B_{0}+k\,t,\label{Bzt}
\end{equation}
where $k$ denotes the scan rate. At time $t$, the system has evolved
into
\begin{equation}
|\psi(t)\rangle=a_{0}|\psi_{g}(t)\rangle+a_{1}|\psi_{e,1}(t)\rangle+a_{2}|\psi_{e,2}(t)\rangle,\label{stateevo2}
\end{equation}
written in the eigenstates of the instantaneous Hamiltonian. The
populations $|a_{i}|^{2}$ depend on the rate at which the critical
points are traversed.  The propagator that converts
$|\psi_{g}(0)\rangle$ into $|\psi(t)\rangle$ can be represented as
\begin{equation}
U(t)=\mathcal{T}\exp[-i\int_{0}^{t}dt'\,\mathcal{H}(t')\label{TorderU}
\end{equation}
where $\mathcal{T}$ denotes the time-ordering operator.

 The dynamics of this system can also be simulated numerically.
For these simulations, as well as for experimental simulations, we
replaced the continuous scan by a stepwise constant effective Hamiltonian
by dividing the total evolution period into $N$ segments with duration
$\delta$. In the limit $N\rightarrow\infty$ this becomes the ideal
case. The evolution propagator is then
\begin{equation}
U(t)\approx\prod_{m=0}^{j}u_{m}=\prod_{m=0}^{j}e^{-i\delta\mathcal{H}(m\delta)},\label{SUt}
\end{equation}
where $t=j\delta$ with $j=0,1..N$.

The initial state should be the ground state for $B_{0}\rightarrow-\infty$,
i.e. the state $|00\rangle$. In the numerical simulations, we chose
$B_{0}=-2$, where the overlap $\langle00|\psi_{g}(0)\rangle|^{2}>0.995$.
Figure \ref{coefficients}(c) shows the populations $|a_{i}|^{2}$
during the scan for $B_{x}=0.1$ and scan rates $k=1$ and $1/20$.
At the beginning of the scan, when the gap between the ground and
first excited states is large, the system can adjust to the change
of the control parameter, and the evolution is therefore adiabatic.
However, when the control parameter approaches the critical point,
where the eigenstates change rapidly, the evolution becomes non-adiabatic
\cite{theoyadi1,Peng05} and the excited states become populated.

The characteristic time, at which the system can adjust to the change
of the eigenstates, is the relaxation time $\tau$. It can be defined
by the inverse of the gap (in frequency units) between the ground
and first excited states. Figure \ref{coefficients}(b) shows the
dependence of $\tau$ on the control field as the system goes through
the phase transition region. It reaches a maximum at the two critical
points in which the ground state is affected.

The non-equilibrium dynamics of this system can be put into the context
of quantum critical systems \cite{Damski05}. Since the two critical
points at $B_{z}^{c}=\pm1$ are fully equivalent, we concentrate on
the critical point at $B_{z}^{c}=-1$, i.e., $B_{z}<0$. If the system
is initially in the ground state, the defect density $D$ corresponds
to the total population of the excited states,
\begin{equation}
D(t)=1-|\langle\psi_{g}(t)|\psi(t)\rangle|^{2}.\label{defect}
\end{equation}

\section{Experimental protocol}

As an experimentally accessible system for testing these predictions,
we chose an NMR quantum simulator. As the quantum register, we used
$^{13}$C-labelled chloroform (CHCl$_{3}$) dissolved in d6-acetone.
The proton and carbon nuclei are assigned as qubits $1$ and $2$.
Data were taken with a Bruker DRX 700 MHz spectrometer. In the doubly
rotating frame, the Hamiltonian of the NMR system is
\begin{equation}
\mathcal{H}_{NMR}=-\pi\nu(\sigma_{z}^{1}+\sigma_{z}^{2})+J\pi\sigma_{z}^{1}\sigma_{z}^{2}/2,\label{Ham}
\end{equation}
where $\nu$ denotes the offset, and $J=215$ Hz denotes the coupling
between the two qubits. The pseudo-pure initial state $|00\rangle$
was prepared by spatial averaging \cite{effectivePure}.

To experimentally determine the overlap $F(t)=|\langle\psi_{g}(t)|\psi(t)\rangle|^{2}$
in Eq.\
(\ref{defect}), we can rewrite it as \cite{zhangNew,BenchNJP}
\begin{equation}
F(t)=|\langle00|P{}^{\dag}(t)\,U(t)\,P(0)|00\rangle|^{2},\label{Fidrew}
\end{equation}
where $P(t)$ denotes the transformation $|00\rangle$ $\to$ $|\psi_{g}(t)\rangle$.

 To optimize the experimental implementation, we used
results from numerical simulations to choose experimental parameters
and design the experiment protocol. The simulations showed:
\begin{enumerate}
\item When $B_{x}$ was chosen as $0.1$ or $0.2$ ( in units of $J\pi/2$),
and $k=1$, $1/2$, $1/3$ or $1/4$ {[}in units of $(J\pi/2)^{2}${]},
the results for the step sizes of the control field $\delta_{B}=0.1$,
$0.04$ and $0.02$ ( in units of $J\pi/2$) are almost identical,
except in the region near $B_{z}=0$, which we exclude from the discussion.
In the experimental implementation, we therefore chose the step size
as $\delta_{B}=0.1$.
\item The propagator $u_{m}$ in Eq. (\ref{SUt}) can be approximated as
\begin{equation}
u_{m}\approx e^{-i\delta B_{x}(\sigma_{x}^{1}+\sigma_{x}^{2})}e^{-i\delta[B_{z}(m\delta)(\sigma_{z}^{1}+\sigma_{z}^{2})+\sigma_{z}^{1}\sigma_{z}^{2}]}.\label{appU}
\end{equation}
Here the discretization results in a reduction of the fidelity from
1 to $>0.994$. The time steps were chosen as $\delta=0.1$, $0.2$,
$0.3$, or $0.4$ {[}in units of $1/(J\pi/2)${]}, for $k=1$, $1/2$,
$1/3$ and $1/4$, respectively.
\item The defect density $D$ remains close to zero ($D<0.005$) for $B_{z}<-1.5$,
i.e. before the system approaches the critical point. We therefore
started the scan at $B_{z}$ = $-1.5$.
\end{enumerate}

\begin{figure}[bt]
\includegraphics[width=9cm]{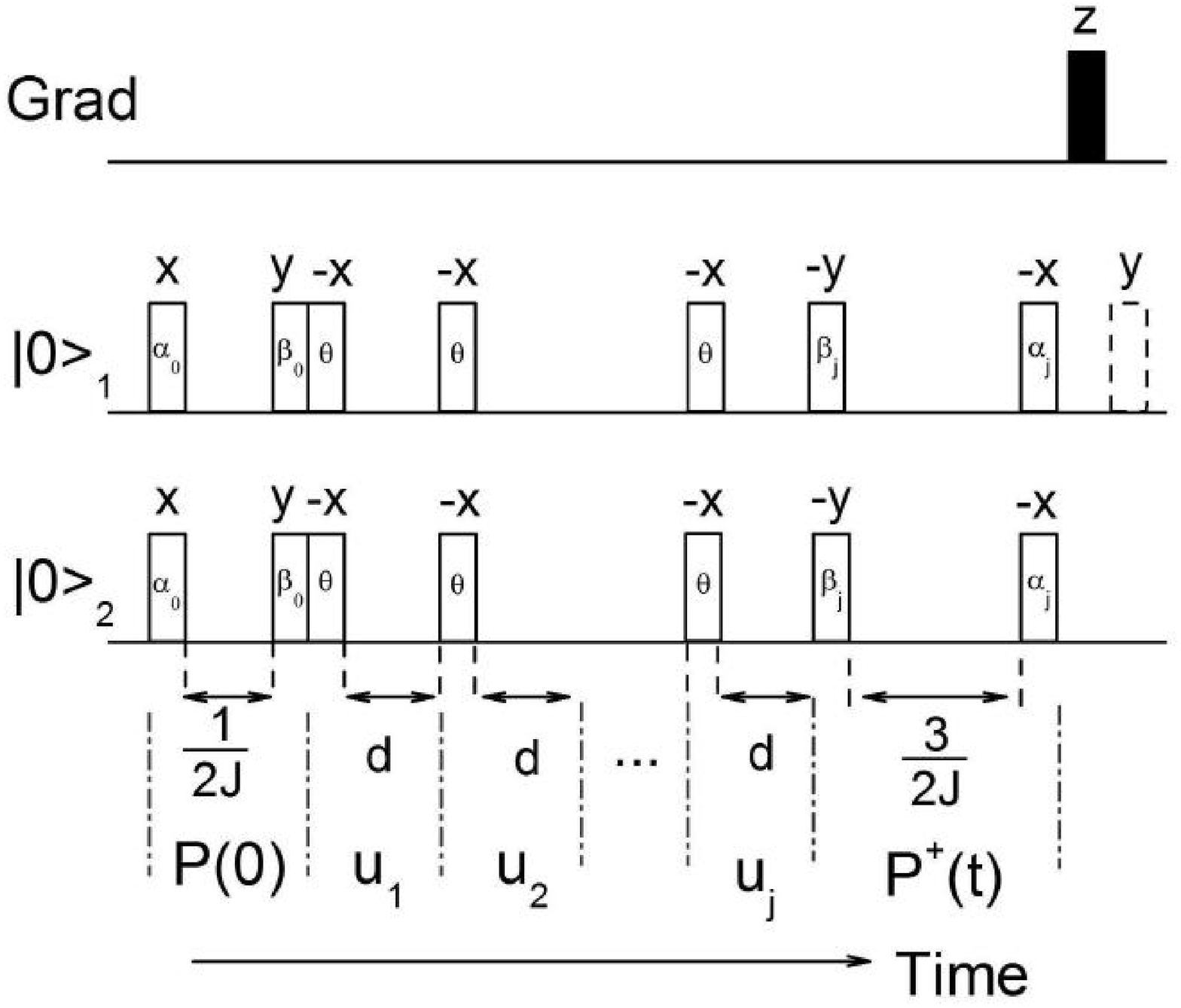} 
\caption{Pulse sequence used for the experimental implementation.
The three horizontal lines from top to bottom denote the gradient,
proton, and carbon channels, respectively. The white rectangles
denote the radio-frequency pulses. Their rotation angles and
rotation axes are shown inside and above the rectangles, the flip
angles $\alpha$ and $\beta$ are given in Eqs.
(\ref{alpha}-\ref{beta}). The flip angles $\theta$ and the delays
$d$ are $\theta=2\delta B_{x}$ and $d=2\delta/(\pi J)$, and the
offsets are $\nu_{m}=(-1.5+0.1m)J/2$ with $m=1,2,...,j$. The black
rectangle denotes a gradient pulse along the z-axis to eliminate
the non-diagonal elements of the density matrix and the dashed
rectangle denotes a $\pi/2$ read out pulse. The durations of the
pulses are $<10\mu s$, short compared to the delays between them.
}

\label{pulseq}
\end{figure}

Figure \ref{pulseq} shows the pulse sequence for the experimental
implementation. We divided the whole scan period into $15$ segments
with identical durations. The corresponding values of the control
parameter $B_{z}$ are $B_{z}(j\delta)=-1.5+0.1j$. We first prepared
the initial state $P(0)|00\rangle$. For given fields $B_{x}$ and
$B_{z}(t)$, the operator
\begin{equation}
P(t)=e^{i\beta(\sigma_{y}^{1}+\sigma_{y}^{2})/2}e^{-i(\pi/4)\sigma_{z}^{1}\sigma_{z}^{2}}e^{i\alpha(\sigma_{x}^{1}+\sigma_{x}^{2})/2}\label{up}
\end{equation}
generates the ground state $|\psi_{g}(B_{x},B_{z}(t)\rangle$ from
state $|00\rangle$, where
\begin{eqnarray}
\cos\alpha & = & c_{0}+c_{1},\label{alpha}
\end{eqnarray}
\begin{eqnarray}
\sin(\beta+\gamma) & = & -\sqrt{2}c_{+}/\sqrt{2-(c_{0}+c_{1})^{2}},\label{beta}
\end{eqnarray}
and $\tan\gamma=\sqrt{1-(c_{0}+c_{1})^{2}}$, The populations $c_{0}$,
$c_{+}$, $c_{1}$ are given in Eq. (\ref{ground}). Figure \ref{pulseq}
shows the pulse sequence used for generating $P(0)$, $U(t)$, and
$P(t)^{\dag}$ .

The system was then allowed to evolve into the state $U(t)P(0)|00\rangle$
and the transformation $P(t)^{\dag}$ was applied, which converts
the ground state of the final Hamiltonian into the $|00\rangle$ state.
Then we applied a gradient pulse to eliminate coherence in this state
since we only need diagonal terms. To obtain the overlap $F(t)$,
we performed partial quantum state tomography \cite{tomo}, using
one $\pi/2$ read out pulse on qubit 1 and invoking the permutation
symmetry of the two qubits.

\begin{figure}[bt]
\centering{}\includegraphics[width=0.99\columnwidth]{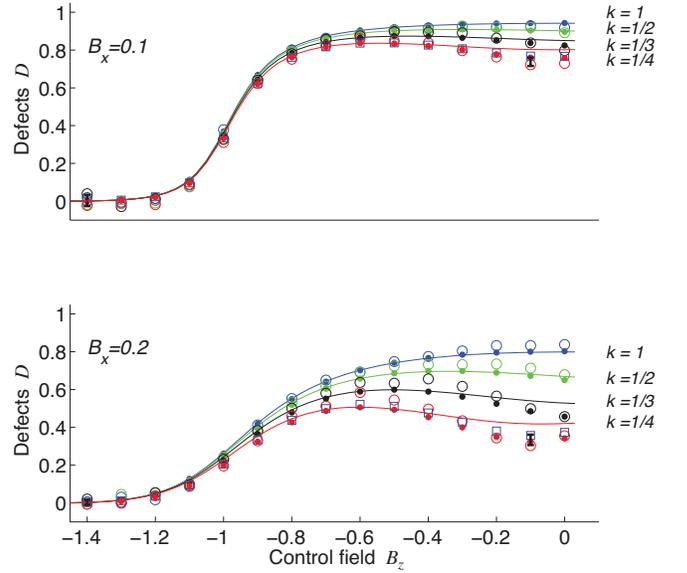}
\protect\protect\protect\protect\caption{Experimental results for
various sweep rates $k$ and transverse fields $B_{x}$, as
indicated in the panel. For each case, the experimental and
simulated data, and the theoretical prediction are marked by
circles, dots and solid curves, respectively. We scanned the
control field from large to small negative values. With increasing
scan duration, the number of required unitary operations
increases, resulting in accumulation of errors and reduction of
the number of defects compared to the ideal curve. The error bars
show the standard deviation of the data points. The squares
represent the results of simulations that include the effect of
transverse ($T_{2}$) relaxation for the sweep rate $k=1/4$. The
$T_{2}$ values for the proton and carbon spins were chosen as 2
and 0.2 s, respectively \cite{tomo}, while the dots correspond to
$T_{2}\rightarrow\infty$. \label{fig:Experimental-results-for}}
\end{figure}

\section{Experimental results and discussion}

We performed the experiment for different values of the transverse
field, $B_{x}=0.1$ and $0.2$ and scan rates $k=1$, $1/2$, $1/3$
and $1/4$. Figure \ref{fig:Experimental-results-for} shows the
results as circles. The experimental results are complemented by
two different types of simulations: the simulation of the ideal
model is shown as a set of curves, while a simulation of the
discretized experiment, using the actual experimental parameters
but ideal pulses is shown as dots. The differences between the
simulated experiment and the theoretical curves reflect the
approximations made by generating the stepwise constant propagator
{[}see Eq.(\ref{appU}){]}. Experimental and simulated data agree
within the experimental uncertainties. Earlier benchmarking
experiments \cite{BenchNJP} showed that the pulse control error is
negligibly small. The differences between theory and experiment
appear to be dominated by transverse relaxation ($T_{2}$) effects.
To estimate the effect of relaxation on the data, we performed
numerical simulations for the case of the slowest scan rate
($k=1/4$), using the experimentally determined relaxation times.
The resulting defect densities are shown as blue squares in Figure
\ref{fig:Experimental-results-for}.

\begin{figure}[bt]
\centering{}\includegraphics[width=0.99\columnwidth]{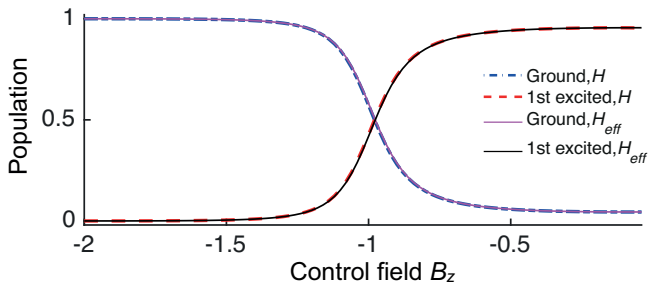}
\caption{ Comparison of the results by numerical simulation
obtained from $\mathcal{H}$ and $\mathcal{H}_{eff}$, with
$B_{x}=0.1$ and $k=1$.}

\label{peff}
\end{figure}

\begin{figure}[bt]
\centering{}\includegraphics[width=0.99\columnwidth]{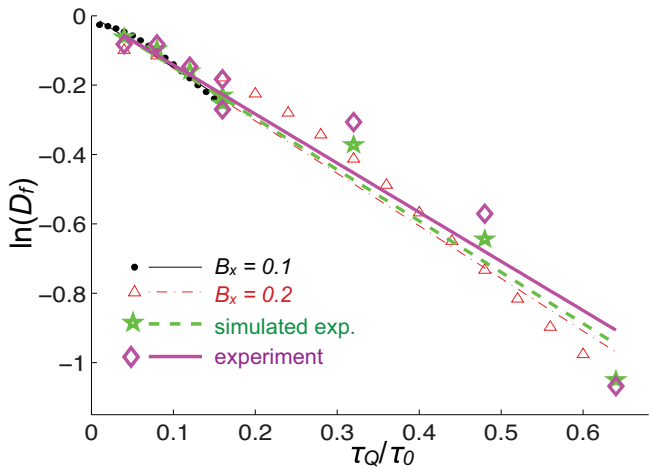}
\protect\protect\protect\protect\caption{Results for estimating
the factor $\alpha$ in Eq. (\ref{freezequ}) by numerical
simulation and experiment, where $D_{f}$ denotes the defects at
$B_{z}=-0.2$.  The data marked by dots and triangles were obtained
by numerically simulating the ideal theoretical model, while the
data marked by diamonds show the experimental results. In the
numerical simulation model, we varied the scan rate $k$ for two
different values of the transverse field, $B_{x}=0.1$ and
$B_{x}=0.2$. The results of a numerical simulation of the
experimental implementation with ideal pulses are shown as stars.
\label{fig:Results-for-estimating}}
\end{figure}

For a simplified description of the anticrossing region, we reduce
the system to an effective two-level system described by the Hamiltonian
\begin{equation}
\mathcal{H}_{eff}(t)=[B_{z}(t)+1]\sigma_{z}+\sqrt{2}B_{x}\sigma_{x}.\label{Heff}
\end{equation}
Here, we have shifted the origin of the energy axis to the center
of the gap. This model (\ref{Heff}) is equivalent to the LZ model,
which was used to simulate the KZM by Damski \cite{Damski05}, up
to a shift of $B_{z}(t)$. The quench time scale $\tau_{Q}$ of the
KZM is proportional to the ratio of the transverse field and the scan
rate: $\tau_{Q}=\sqrt{2}B_{x}/k$ \cite{PhysRep98,Damski05}. In Figure
\ref{peff}, we show the good agreement between the results by numerical
simulation of the dynamics of $\mathcal{H}_{eff}$ and $\mathcal{H}$
in our region of interest.

Since we are mostly interested in the behavior near the critical point
$B_{z}^{c}=-1$, we rescale the distance from the critical point by
dividing the time $|B_{z}+1|/k$ required to reach $B_{z}^{c}$ by
$\tau_{Q}$: $\varepsilon=|B_{z}(t)+1|/(k\,\tau_{Q})$. In these units,
the relaxation time becomes
\begin{equation}
\tau(t)=\frac{\tau_{0}}{\sqrt{1+\varepsilon^{2}}}\label{relaxtre}
\end{equation}
with the maximum of the relaxation time $\tau_{0}=1/(2\sqrt{2}B_{x})$.

According to the KZM model, the boundary between the adiabatic and
non-adiabatic ( or impulse \cite{Damski05}) regimes is given by the
freeze-out time $\hat{t}$, which is related to the relaxation time
$\tau$ by the equation \cite{Zurek,Damski05}
\begin{equation}
\tau(\hat{t})=\alpha\hat{t}\label{freezequ}
\end{equation}
where $\alpha=\mathcal{O}(1)$ and independent of $\tau_{Q}$ and
$\tau_{0}$. The freeze-out time $\hat{t}$ indicates the distance
to the critical point. The constant $\alpha$ is also related to the
final defect density $D_{f}$ after the passage through the critical
point
\begin{equation}
D_{f}\approx e^{-\alpha\tau_{Q}/\tau_{0}}=e^{-\alpha4B_{x}^{2}/k}.\label{Dsequ}
\end{equation}
 The derivation of this relation is presented in the
appendix.

Since we are only interested here in the effect of the passage
through the first critical point, we determined the 'final' defect
density $D_{f}$ at $B_{z}=-0.2$, well after the first critical
point, but before the second critical point. To test the scaling
relation \eqref{Dsequ}, we compare defect densities for a range of
scan rates $k$ and two transverse field strengths $B_{x}=0.1$ and
$0.2$. Figure \ref{fig:Results-for-estimating} shows the
dependence of $D_{f}$ on $\tau_{Q}/\tau_{0}$ (or the scan duration
scaled as $B_{0}/(4B_{x}^{2})$ ) for two sets of simulated data
and for the experimental data of Figure
\ref{fig:Experimental-results-for}.  Linear fits of the simulated
and experimental data sets yield the parameters listed in Table
\ref{tablefitp}. These results agree well with the theoretical
predictions of Eq. (\ref{Dsequ}) and $\alpha=\mathcal{O}(1)$, as
well as with the LZ-formula, which predicts $\alpha$ = $\pi/2$
\cite{LZForm,PhysRep98}. The difference between simulated and
actual experiments is mostly due to transverse relaxation.

\begin{table}
\begin{tabular}{|c|c|c|c|c|}
\hline
 & simulation  & simulation  & simulated  & experiment \tabularnewline
 & $B_{x}=0.1$  & $B_{x}=0.2$  & experiment  & \tabularnewline
\hline
$\alpha$  & 1.49  & 1.51  & 1.48  & 1.42 \tabularnewline
\hline
$r$  & 0.99  & 0.99  & 0.98  & 0.96 \tabularnewline
\hline
\end{tabular}\protect\caption{Parameters obtained by fitting the data from the simulation
and the experiment. $r$ denotes the correlation coefficient of the
linear fit.}

\label{tablefitp}
\end{table}

\section{Entanglement from the non-equilibrium phase transition}

\begin{figure}[h]
\centering{}\includegraphics[width=0.99\columnwidth]{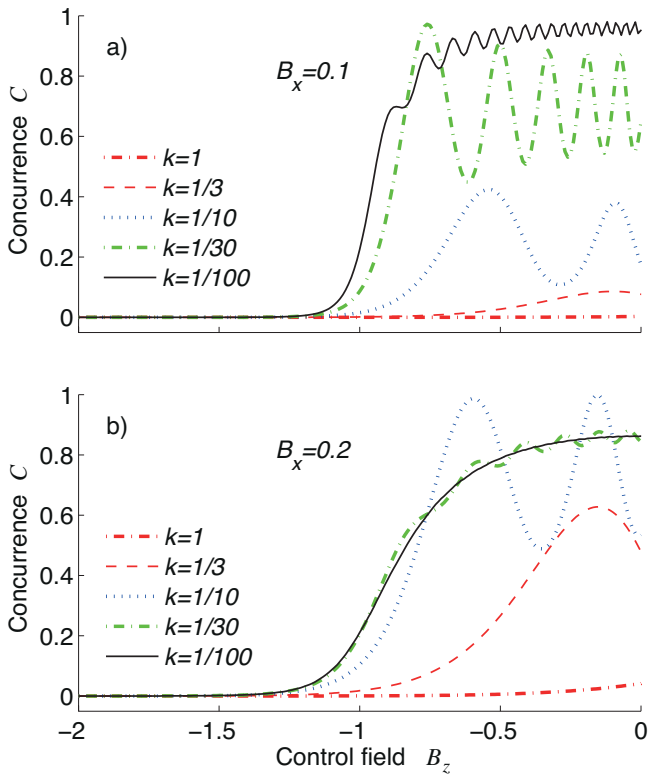}
\protect\protect\protect\protect\protect\protect\caption{
Concurrence, as a measure of entanglement, as a function of the
control parameter $B_{z}$, for different values of the transverse
field $B_{x}$ and scan rate $k$. \label{fig:Concurrence,-as-a}}
\end{figure}

As a function of the control parameter, the two-qubit system used
for these quantum simulations has different ground states; for
$|B_{z}|>1$, the ground state is separable, for $|B_{z}|<1$ it is
entangled, see Eq. (\ref{groundBx0}). Accordingly, we prepared the
system in a separable state, but depending on the adiabaticity,
the passage through the critical points creates entanglement in
the system. This entanglement generation can be controlled by the
transverse field and the scan rate of the control field
\cite{note}.  Figure \ref{fig:Concurrence,-as-a} shows the results
of numerical simulations, using the concurrence \cite{concurr} as
the entanglement measure. When $B_{x}$ is small and the scan is
fast, such as in the case of the red dot-dashed curve in figure
(a), the system remains in the separable state and almost no
entanglement is generated, corresponding to many defects. For
larger $B_{x}$ and slower scan rate, the probability that the
system remains in the ground state increases and the two qubits
become entangled during the passage through the critical point,
while the defect density decreases.  In the limit of a slow scan,
the evolution becomes adiabatic, and the entanglement becomes a
measure of the QPT \cite{Peng05}.  At intermediate scan rates, the
concurrence shows significant oscillations, indicating that the
system is far from equilibrium, in a superposition of the two
lowest eigentstates. The lower sweep rates require longer scan
times, e.g., about 30 and 90 ms for $k=1/10$, and $1/30$,
respectively. In such cases, the errors from the relaxation
effects increase. One possible solution for exploring this regime
would be to choose a system with stronger couplings, such as
dipolar coupled spins in liquid crystal solvents
\cite{ChuangLC99}.

\section{Conclusion}

We have experimentally simulated the KZM of defect generation in an
interacting quantum system consisting of a pair of coupled nuclear
spins. We controlled the spins by applying near-resonant radio-frequency
fields. Our model contains several critical points; here, we focused
on the one that is encountered first as the control parameter is swept
from large negative to large positive fields. We demonstrated the
formation of defects during the evolution starting from the ground
state of the initial Hamiltonian. Like in the LZ model, the relaxation
time of the system passes through a maximum at the critical point,
but does not diverge, in contrast to the original KZM. We also used
our model system to verify the validity of the Zurek equation and
determine the freeze-out time.

Our work can be considered as a first step of a more general
strategy to study the KZM in interacting many body systems with
QPTs. While we used a comparatively simple two-qubit system, it
can be readily extended to larger systems, which allow to simulate
more complex quantum critical phenomena, such as in the quantum
Ising model \cite{Sachdev} or the generation of entanglement in
the non-equilibrium phase transition.
\begin{acknowledgments}
This work was supported by the DFG through grant grant 192/19-2, the
DAAD through grant 57052334, and the government of Canada.
\end{acknowledgments}

\section*{Appendix: Calculation of the final defect density $D_{f}$}

As discussed above, the freeze-out time $\hat{t}$ quantifies the
distance to the critical point. For the following, we redefine the
origin of the time axis as $t\rightarrow t+\frac{B_{0}+1}{k}$. With
this change, the effective Hamiltonian (\ref{Heff}) has exactly the
LZ form,
\begin{equation}
H_{eff}=kt\sigma_{z}+\sqrt{2}B_{x}\sigma_{x}.\label{Heff2LZ}
\end{equation}
Therefore we can compute the final defect density $D_{f}$ after the
passage through the critical point by following Damski \cite{Damski05}.
In particular, if the initial state at time $t_{0}$ is
\begin{equation}
\left|\psi(t_{0})\right>=\left|\psi_{g}(t_{0})\right>,\label{stateini2}
\end{equation}
with $\left|\psi_{g}(t)\right>$ and $\left|\psi_{e}(t)\right>$ the
ground and excited states at time $t$, we use the KZM assumptions
to estimate the evolution of the wave function after the first adiabatic
regime,
\begin{equation}
\left|\psi(-\hat{t})\right>\simeq\left|\psi_{g}(-\hat{t})\right>,\label{adia1}
\end{equation}
after the impulse regime,
\begin{eqnarray}
\left|\psi(\hat{t})\right> & \simeq & \left|\psi(-\hat{t})\right>\simeq\left|\psi_{g}(-\hat{t})\right>\nonumber \\
 & = & \left<\psi_{g}(\hat{t})|\psi_{g}(-\hat{t})\right>\left|\psi_{g}(\hat{t})\right>\nonumber \\
 & + & \left<\psi_{e}(\hat{t})|\psi_{g}(-\hat{t})\right>\left|\psi_{e}(\hat{t})\right>,\label{impulse1}
\end{eqnarray}
and at the final time $t_{f}$ after second adiabatic regime,
\begin{eqnarray}
\left|\psi(t_{f})\right> & \simeq & \left<\psi_{g}(\hat{t})|\psi_{g}(-\hat{t})\right>\left|\psi_{g}(t_{f})\right>\nonumber \\
 & + & \left<\psi_{e}(\hat{t})|\psi_{g}(-\hat{t})\right>\left|\psi_{e}(t_{f})\right>.\label{impulse2}
\end{eqnarray}
From Eqs. (\ref{impulse1}) to (\ref{impulse2}), we use the fact
that in the adiabatic evolution from $\hat{t}$ to $t_{f}$, the coefficients
before $\left|\psi_{g}(\hat{t})\right>$ and $\left|\psi_{e}(\hat{t})\right>$
remain unchanged, while the instantaneous eigenstates evolve to $\left|\psi_{g}(t_{f})\right>$
and $\left|\psi_{e}(t_{f})\right>$. Therefore, the KZM allows us
to approximate $D_{f}$
\begin{eqnarray*}
D_{f} & = & \left|\left<\psi(t_{f})|\psi_{e}(t_{f})\right>\right|^{2}\\
 & \simeq & \left|\left<\psi_{e}(\hat{t})|\psi_{g}(-\hat{t})\right>\right|^{2}.
\end{eqnarray*}
Since $|\psi_{e}(\hat{t})\rangle$ and $|\psi_{g}(-\hat{t})\rangle$
are instantaneous eigenstates of $H_{eff}$ in (\ref{Heff2LZ}),
\begin{equation}
\left(\begin{array}{c}
|\psi_{g}(t)\rangle\\
|\psi_{e}(t)\rangle
\end{array}\right)=\left(\begin{array}{cc}
\cos(\theta/2) & \sin(\theta/2)\\
-\sin(\theta/2) & \cos(\theta/2)
\end{array}\right)\left(\begin{array}{c}
|0\rangle\\
|1\rangle
\end{array}\right),
\end{equation}
where $\sin(\theta)=1/\sqrt{1+\epsilon^{2}}$, $\cos(\theta)=\epsilon/\sqrt{1+\epsilon^{2}}$,
and $\epsilon=kt/\sqrt{2}B_{x}$. Using this, the final defect density
is
\begin{equation}
D_{f}\simeq\frac{\hat{\epsilon}}{1+\hat{\epsilon}^{2}},\label{Dfin2}
\end{equation}
where $\hat{\epsilon}\equiv\epsilon(\hat{t})$ is the solution of
Eq. (18) {[}15{]},
\begin{equation}
\hat{\epsilon}\equiv\epsilon(\hat{t})=\frac{1}{\sqrt{2}}\sqrt{\sqrt{1+\frac{4}{x_{\alpha}^{2}}}-1},\ \ \ x_{\alpha}=\alpha\frac{\tau_{Q}}{\tau_{0}}.
\end{equation}
Substituting this into expression \eqref{Dfin2} and expanding for
fast transitions ($\tau_{Q}\rightarrow0$ with fixed $\tau_{0}$),
we obtain Eq. (\ref{Dsequ}).



\begin{thebibliography}{10}
\bibitem{note} In the current experiment, which uses a pseudo-pure
initial state, this corresponds to pseudo-entanglement.
\end{thebibliography}
\end{document}